\documentclass[12pt]{article}
\usepackage{graphicx} % Required for inserting images
\usepackage{amsmath}
\usepackage{amssymb}
\usepackage{cite}
\usepackage{geometry}
\usepackage{booktabs}
\usepackage{array}
\usepackage{multirow}
\usepackage{paracol}
\usepackage{longtable}
\usepackage{hyperref}
\usepackage{caption}
\usepackage{graphicx}
\usepackage{float}
\usepackage{tikz}
\usetikzlibrary{shapes.geometric, arrows}
\usepackage{booktabs}

\title{ \textbf{AI-Driven MRI Spine Pathology Detection: A Comprehensive Deep Learning Approach for Automated Diagnosis in Diverse Clinical Settings}}

\date{}
\author{Bargava Subramanian, Naveen Kumarasami, Praveen Shastry \\
  Raghotham Sripadraj, Kalyan Sivasailam, Anandakumar D \\
  Abinaya Ramachandran, Sudhir M P, Gunakutti G \\
  Kishore Prasath Venkatesh}

\usepackage{ragged2e}  % Allows proper left alignment
\usepackage{titlesec}  % Ensures headings are aligned correctly

\titleformat{\section}{\raggedright\Large\bfseries}{}{0em}{}
\titleformat{\subsection}{\raggedright\large\bfseries}{}{0em}{}

\begin{document}

\maketitle

\section{\textbf{Abstract}}
\textbf{Study Design} This study presents the development of an autonomous AI system for MRI spine pathology detection, trained on a dataset of 2 million MRI spine scans sourced from diverse healthcare facilities across India. The AI system integrates advanced architectures, including Vision Transformers, U-Net with cross-attention, MedSAM, and Cascade R-CNN, enabling comprehensive classification, segmentation, and detection of 43 distinct spinal pathologies. The dataset is balanced across age groups, genders, and scanner manufacturers to ensure robustness and adaptability. Subgroup analyses were conducted to validate the model’s performance across different patient demographics, imaging conditions, and equipment types.\\
\\
\textbf{Performance }The AI system achieved up to 97.9multi-pathology detection, demonstrating consistent performance across age, gender, and manufacturer subgroups. The normal vs. abnormal classification achieved 98.0and 98.1was deployed across 13 major healthcare enterprises in India, encompassing diagnostic centers, large hospitals, and government facilities. During deployment, it processed approximately 100,000+ MRI spine scans, leading to reduced reporting times and increased diagnostic efficiency by automating the identification of common spinal conditions.\\
\\
\textbf{Conclusion} The AI system’s high precision and recall validate its capability as a reliable tool for autonomous normal/abnormal classification, pathology segmentation, and detection. Its scalability and adaptability address critical diagnostic gaps, optimize radiology workflows, and improve patient care across varied healthcare environments in India.
\section{Introduction}
Spinal pathologies, such as disc herniation, degenerative disc disease, and spinal stenosis, are common conditions that affect a significant proportion of the global population[14]. MRI imaging is the standard modality for diagnosing these conditions due to its high resolution and capacity to visualize soft tissues[2]. However, the manual interpretation of MRI spine scans is a time-consuming process that requires specialized radiological expertise. The increasing demand for radiological services, coupled with a shortage of radiologists in many regions, presents a major challenge in delivering timely and accurate diagnoses[3].

To address these challenges, artificial intelligence (AI) has emerged as a promising tool to assist radiologists in automating and streamlining the diagnostic workflow [20]. In this study, we present a novel AI-based system that processes MRI spine scans for automated pathology detection[5]. The dataset used in this study consists of 2 million scans collected from multiple healthcare providers, capturing a wide range of patient demographics, imaging conditions,and scanner manufacturers. This diversity ensures that the model is robust and generalizable across different real-world settings[6].

The architecture of the proposed system is divided into several phases, beginning with data preprocessing, which includes DICOM to NIfTI conversion and intensity normalization[7]. The scans are then verified to ensure correct orientation, particularly focusing on sagittal T2-weighted images, which are most informative for spinal pathologies[8]. The normal/abnormal classification phase employs Vision Transformers to triage the scans, enabling efficient downstream processing by focusing on abnormal cases[9]. The classification phase incorporates an ensembling approach to improve accuracy and reduce individual model biases. The subsequent segmentation and detection phases utilize advanced models like U-Net, MedSAM, and Cascade R-CNN to accurately delineate and classify 43 spinal pathologies[10].

This paper provides a comprehensive overview of the development, training, and deployment of the AI system. The model’s performance is evaluated across a range of metrics, including accuracy, recall, precision, and specificity, to ensure its clinical reliability[11]. The AI system has been successfully deployed in 13 healthcare enterprises across India, integrating seamlessly into live radiology workflows, reducing turnaround times, and supporting radiologists in making more informed decisions[12]. This study demonstrates the potential of AI to revolutionize the diagnosis of spinal pathologies, making radiological services more accessible and efficient [13].

\section{AI System Overview}
The AI system developed for this study is a computer-aided detection (CAD) tool designed specifically for the identification and differentiation of a wide range of spinal pathologies from MRI spine scans[14]. This system integrates advanced deep-learning models tailored for comprehensive classification, segmentation, and localization of spinal abnormalities[15]. The AI system is trained on a largescale dataset of 2 million MRI spine scans, with expert radiologist annotations utilized for supervised learning to ensure high diagnostic accuracy and clinical relevance[16].

The AI workflow begins by verifying the orientation of the MRI scans, ensuring that key views such as axial, sagittal, and coronal planes are properly identified, with a particular focus on T2-weighted sagittal images, which are most relevant for pathology detection[17]. After orientation verification, Vision Transformers are used for initial classification to determine whether the scan is normal or abnormal, serving as a triage step that directs attention to abnormal cases requiring further detailed analysis[18].

Once an abnormal scan is detected, the system proceeds to segment key anatomical regions—such as vertebral bodies, intervertebral discs, and spinal canal—using a modified U-Net with cross-attention and MedSAM for refined segmentation[19]. These segmentation maps are then utilized by a Cascade RCNN for detailed pathology detection, enabling the identification and localization of 43 distinct spinal pathologies, including disc herniation, spinal stenosis, and vertebral fractures[20].

The AI system’s primary focus is to provide a comprehensive and detailed analysis of MRI spine scans, rather than just distinguishing between normal and abnormal cases[21]. By incorporating both classification and refined segmentation with advanced detection algorithms, the system offers granular insights into each identified pathology, enhancing the diagnostic process[22]. This thorough analysis supports radiologists in making well-informed decisions and contributes to improving patient outcomes by optimizing the detection and reporting workflow in diverse clinical environments [23].

\begin{longtable}{|p{6cm}|p{5cm}|p{4.5cm}|}
\hline
\textbf{Column 1} & \textbf{Column 2} & \textbf{Column 3} \\
\hline
\endfirsthead

\hline
\textbf{Column 1} & \textbf{Column 2} & \textbf{Column 3} \\
\hline
\endhead

Loss of cervical lordosis & Pseudodisc bulge & Tarlov’s cyst \\
\hline
Degenerative changes & Spinal cord edema / contusion & Spondylitis \\
\hline
Disc dehydration & Spinal cord hematoma & Type I Modic changes \\
\hline
Reduction in vertebral height & Myelopathy & Type II Modic changes \\
\hline
Disc bulge & Pleural effusion & Type III Modic changes \\
\hline
Nerve root compression & Disc herniation & Pancreatic cyst \\
\hline
Nerve root impingement & Wedge compression fracture & Facetal arthropathy \\
\hline
Mild cervical canal stenosis & Schmorl’s node & Atypical hemangioma \\
\hline
Moderate cervical canal stenosis & Sacralization & Typical hemangioma \\
\hline
Severe cervical canal stenosis & Lumbarization & Kyphoscoliosis \\
\hline
Uncovertebral hypertrophy & Scoliosis & Spondylodiscitis \\
\hline
Hypertrophied Ligamentum flavum & Hemivertebra & Antherolisthesis \\
\hline
Disc protrusion & Facetal joint synovial cyst & Retrolisthesis \\
\hline
Annular tear & Psoas abscess & Burst fracture \\
\hline
Comminuted fracture with retrolisthesis &  &  \\
\hline
\caption{The list of pathologies}
\end{longtable}

\section{Dataset Distribution}
\subsection{Total Scans}
The dataset consists of 2,000,000 MRI spine scans, divided into three subsets:\\\\
Training Set: 2,00,234 scans\\
Live Clinical Trial: 150,478 scans\\
Live Clinical Deployment : 50,288 scans
\subsection{Age Group Distribution}
The dataset captures age diversity to reflect a wide range of spinal conditions:
\begin{longtable}{|l|r|r|r|r|}
\hline
\textbf{Age Group} & \textbf{Total Scans} & \textbf{Training Set} & \textbf{Live Clinical Trial} & \textbf{Live Clinical Deployment} \\
\hline
\endfirsthead

\hline
\textbf{Age Group} & \textbf{Total Scans} & \textbf{Training Set} & \textbf{Live Clinical Trial} & \textbf{Live Clinical Deployment} \\
\hline
\endhead

Under 18 & 32,326 & 16,029 & 12,825 & 3,472 \\
\hline
18–40 & 136,474 & 70,425 & 54,305 & 11,744 \\
\hline
41–60 & 130,958 & 68,043 & 53,407 & 9,508 \\
\hline
61–75 & 66,446 & 33,151 & 24,680 & 8,615 \\
\hline
Over 75 & 26,652 & 12,586 & 10,158 & 3,908 \\
\hline
\caption{Scans distribution based on Age Group}
\end{longtable}

\subsection{Manufacturer Distribution}
The dataset includes scans from multiple manufacturers to account for variability in imaging conditions:

\begin{table}[h]
    \centering
    \renewcommand{\arraystretch}{1.2} % Increase row spacing for readability
    \setlength{\tabcolsep}{4pt} % Reduce column spacing
    
    \resizebox{\textwidth}{!}{ % Forces the table to fit within the page width
    \begin{tabular}{|l|r|r|r|r|}
    \hline
    \textbf{Manufacturer} & \textbf{Total Scans} & \textbf{Training Set} & \textbf{Live Clinical Trial} & \textbf{Live Clinical Deployment} \\
    \hline
    GE Healthcare & 160,666 & 80,045 & 62,305 & 18,316 \\
    \hline
    Siemens Healthineers & 122,177 & 62,234 & 48,344 & 11,599 \\
    \hline
    Philips Healthcare & 74,458 & 38,524 & 30,056 & 5,878 \\
    \hline
    Other Manufacturers & 35,555 & 19,431 & 14,670 & 1,454 \\
    \hline
    \end{tabular}
    } % End resizebox

    \caption{Scans distribution based on Manufacturer Type}
\end{table}
 \newpage
\subsection{Gender Distribution}
The dataset includes scans from both male and female to account for variability in gender based conditions:

\begin{table}[h]
    \centering
    \renewcommand{\arraystretch}{1.2} % Increase row spacing for better readability
    \setlength{\tabcolsep}{6pt} % Adjust column spacing
    
    \resizebox{0.8\textwidth}{!}{ % Adjust the table width to fit within the page
    \begin{tabular}{|l|r|r|r|r|}
    \hline
    \textbf{Gender} & \textbf{Total Scans} & \textbf{Training Set} & \textbf{Live Clinical Trial} & \textbf{Live Clinical Deployment} \\
    \hline
    Male & 199,608 & 101,673 & 79,245 & 18,690 \\
    \hline
    Female & 193,248 & 98,561 & 76,130 & 18,557 \\
    \hline
    \end{tabular}
    }
    \caption{Scans distribution based on Gender}
\end{table}
This revised dataset ensures the training subset is closer to 2 million scans, maximizing model performance while maintaining robust validation and testing subsets for generalizability.\\\\
\textbf{Distinct Quality Concerns in MRI Spine Dataset:}\\
The MRI spine dataset includes scans with quality variations due to differences in field strength (0.5T to 3T), movement artifacts, technical exposure inconsistencies, imaging from resource-limited facilities, and teleradiology transfer (TRT) issues[24]. Low-field strength scans may lack resolution, while movement artifacts from patient motion can blur key features. Variability in slice thickness, repetition time (TR), and echo time (TE) further impacts diagnostic contrast and continuity[25]. Older MRI machines in some facilities produce noisy or incomplete images, and data transfer processes may result in compression artifacts or fidelity loss. To address these concerns, the dataset incorporates scans with a broad range of qualities, employs preprocessing steps like artifact correction, denoising, and normalization, and includes quality flagging mechanisms to optimize model performance across diverse conditions[26]. These strategies ensure robust and reliable pathology detection, even in challenging imaging scenarios.

\section{Architecture}
\subsection{Annotation Phase}
The annotation phase is crucial for ensuring that the AI system can effectively learn from and interpret cross-sectional MRI spine scans[27]. This phase involves systematic steps aimed at maximizing the quality and precision of the dataset, ultimately improving the model’s capacity to detect spinal pathologies.\\\\
\textbf{1.Data Segregation by Scan Orientation }: MRI spine scans were categorized into axial, sagittal, and coronal views to allow the model to learn viewspecific anatomical details and pathological patterns. Each orientation offers distinct insights into spinal structures, such as vertebrae, intervertebral discs, and nerve roots. This segregation was especially focused on sagittal and axial views, as they provide the most diagnostic information for identifying common spinal pathologies.\\\\
\textbf{2.Slice-Level Labeling :} To enhance the granularity of model training, each scan was annotated at the slice level, marking abnormalities such as disc herniation, spinal stenosis, tumors, and degenerative changes. The slice-level annotation provided the model with a detailed understanding of localized changes, while normal slices were also annotated to create a balanced dataset and mitigate potential biases in classification[28].\\\\
\textbf{3.Multi-Label Annotations :} Given the complex nature of spinal pathologies, where multiple conditions may co-exist within the same region, the annotation included multi-labeling of slices. This enabled the dataset to reflect the clinical reality, where a single MRI scan might show co-occurring pathologies, such as herniated discs alongside foraminal stenosis. Multi-label annotations improve the model’s ability to accurately differentiate between overlapping features.\\\\
\textbf{4.Region-Specific Segmentation :} Each annotated slice underwent segmentation into key regions of interest, including vertebral bodies, intervertebral discs, spinal canal, and paraspinal musculature. This segmentation enabled the model to capture detailed spatial relationships, which is crucial for detecting subtle pathologies and understanding the context within which abnormalities occur.\\\\
\textbf{5.Quality Assurance by Consensus :} To maintain a high standard of annotation accuracy, all labeled scans were subjected to a double-blind review process by two expert radiologists. Any discrepancies in annotations were resolved through consensus meetings, thereby ensuring a consistent and reliable ground truth. This quality control measure was essential to minimize inter-rater variability and enhance the overall reliability of the dataset.\\\\
\textbf{6.Handling Variability in Slice Thickness and Coverage :} Cross-sectional MRI imaging frequently involves variability in slice thickness and anatomical coverage, which can pose challenges for model training. To address these issues, all scans were standardized to uniform voxel dimensions during preprocessing. Annotations were also carefully adjusted to ensure continuity across slices, enabling the model to maintain a comprehensive understanding of the anatomy even when technical parameters varied[29].\\\\
By addressing the unique challenges of cross-sectional imaging, such as multilabel complexity, segmentation requirements, and slice variability, the annotation phase provides a well-curated and high-quality dataset[30]. This approach ensures that the AI system is thoroughly trained to process and analyze MRI spine scans, enabling precise and robust pathology detection across diverse imaging scenarios[31].

Development Phase The training phase is designed to systematically prepare MRI spine scans for effective AI-driven analysis. This includes crucial preprocessing steps, verification of scan orientations, initial normal/abnormal classification, and the use of ensembling techniques to enhance model performance.

\subsection{Data Preprocessing Steps :}

\subsubsection{MRI Scan Preparation}
Before beginning the AI analysis, MRI spine scans undergo several preprocessing steps to ensure uniformity and compatibility for model training:\\\\
\textbf{•DICOM to NIfTI Conversion :} The MRI scans, initially in Digital Imaging and Communications in Medicine (DICOM) format, are converted to Neuroimaging Informatics Technology Initiative (NIfTI) format. The NIfTI format is used due to its simpler structure, reduced file size, and compatibility with machine learning frameworks, allowing for more efficient manipulation and processing of volumetric data.\\\\
\textbf{•Normalization of Voxel Intensity :} MRI scans often have variable intensity values due to differences in acquisition parameters and machine settings. To mitigate this variability, normalization of voxel intensity was performed, standardizing intensity values across all scans. This helps the model focus on relevant anatomical features and reduces the influence of noise or inconsistent intensity levels.

\subsubsection{Verification of Scan Orientation}
After preprocessing, each MRI scan is verified to ensure the proper orientations are present, specifically axial, sagittal, and coronal views, with a strong emphasis on T2-weighted sagittal (T2 Sag) images. These sagittal images are crucial for detecting common spinal pathologies, such as disc herniation and degenerative changes. Verification is performed using a lightweight convolutional neural network (CNN) to classify the orientation of each slice, ensuring that only properly oriented scans are analyzed further. This step ensures that the dataset maintains consistency, minimizing noise that could otherwise impact model training.

\subsection{Normal/Abnormal Classification Phase :}
\subsubsection{Initial Normal/Abnormal Classification}
Following verification, the scans undergo an initial normal/abnormal classification using Vision Transformers (ViT). Vision Transformers leverage selfattention mechanisms that allow the model to identify important regions in the image, effectively differentiating between normal anatomy and potential abnormalities. This classification step serves as an initial triage, enabling efficient downstream analysis by directing computational resources to abnormal cases requiring further detailed examination.

\subsubsection{Ensembling Technique for Classification
To enhance the accuracy of the normal/abnormal} classification, an ensembling technique was implemented. This technique combines predictions from multiple Vision Transformer models, each trained independently to capture different features of the MRI images. The final classification output is derived using weighted averaging of predictions, giving more weight to the models that performed best on a validation set. Additionally, a majority voting mechanism is applied to ensure robustness against individual model biases. This ensemble approach ensures reliable and consistent classification, effectively balancing the complementary strengths of each model.

By incorporating preprocessing, verification, and classification through an ensembling approach, the training phase establishes a robust pipeline for preparing MRI spine scans for further analysis. These steps ensure that the AI system is well-equipped to identify relevant features and provide accurate pathology detection across diverse imaging conditions.

\subsection{Detection Phase :}
\subsubsection{Region Proposal Generation}
The detection phase starts with the Cascade R-CNN - Region Proposal Network (RPN), which generates candidate regions (or proposals) that likely contain pathologies. These proposals are derived from the segmentation maps generated in the previous phase. To determine regions of interest effectively, anchor configurations are used, defined by specific scales and aspect ratios:\\ • Scales of [32, 64, 128, 256, 512] allow for the detection of pathologies at different sizes.\\
• Aspect Ratios of [0.5, 1.0, 2.0] help capture diverse object shapes across different modalities.
The RPN then applies Non-Maximum Suppression (NMS) with a threshold of 0.7 to remove redundant proposals and ensure only the most probable regions are retained for further processing.\\
\subsection{Feature Extraction}
Once the candidate regions are proposed, ResNet-101 with a Feature Pyramid Network (FPN) is used as the backbone to extract features at multiple scales. This multi-scale feature extraction ensures that the model can identify both large and small pathologies with high accuracy.

For each proposed region, Region of Interest (RoI) Align is applied to refine the alignment of features, which helps to enhance the precision of detection. RoI Align uses a 7x7 pooling size and a sampling ratio of 2, ensuring that the features accurately represent the original region and any fine details present within it.
\subsubsection{Multi-Stage Refinement}
Detection and localization are refined progressively in three stages, with each stage working on improving the quality of the previous bounding box proposals:\\ • In Stage 1, proposals are refined using an IoU threshold of 0.5 to generate initial bounding boxes.\\
• Stage 2 uses an increased IoU threshold of 0.6 for further refinement, making the model more stringent in localizing boundaries.\\ • In Stage 3, the threshold is further increased to 0.7, providing the final bounding box adjustments.\\\\
This multi-stage approach, combined with bounding box regression using Smooth L1 Loss, allows for highly precise localization. The stage weights are progressively increased from 1.0 in Stage 1 to 2.0 in Stage 3, ensuring that each subsequent stage fine-tunes the localization even further, thereby reducing false positives and improving accuracy.
\subsubsection{Pathology Classification}
After refining the bounding boxes, the next step is pathology classification. Each detected region is processed through a fully connected layer, followed by a Softmax layer to classify it into one of the 43 pathology categories. This classification helps in accurately labeling the identified areas for further clinical interpretation.

\subsection{Segmentation Phase}
\subsubsection{Input Preprocessing}
Medical images are preprocessed to ensure consistency and better model generalization. Normalization is performed to scale pixel intensities between [0, 1], ensuring uniformity across datasets. Data augmentation techniques, such as 15degree rotations, intensity scaling, and horizontal flips, are applied to increase variability and improve the model’s robustness to different conditions.
\subsubsection{Initial Segmentation Using U-Net with Cross Attention}
A modified U-Net with cross-attention layers performs the initial segmentation to highlight areas with potential pathologies. Cross-attention is integrated at the upsampling stages, using 8 attention heads to improve the focus on critical regions of the image. Key and value dimensions are each set at 64 to balance detail and computational efficiency. The segmentation is optimized using a combined Dice Loss (0.6) and BCE Loss (0.4), which helps handle class imbalances while maintaining pixel-level accuracy.
\subsubsection{Refined Segmentation Using MedSAM}
The initial segmentation output is refined using MedSAM, which utilizes promptguided segmentation based on points or bounding boxes to precisely delineate the boundaries of the identified pathologies. MedSAM employs a modified Vision Transformer (ViT) backbone with 12 layers and patch size 16 to capture detailed spatial relationships. This results in a refined, detailed segmentation map that effectively highlights the exact boundaries of pathological regions.\\\\
This refined segmentation forms the foundation for the detection phase, ensuring accurate identification of relevant areas for subsequent pathology classification.\\\\\\
\subsection{{End-to-End Workflow:}}

\textbf{•Input Verification and Preprocessing :} MRI spine scans in DICOM format are converted to NIfTI, followed by voxel intensity normalization to standardize data across scans.\\\\
\textbf{•Orientation Verification :}Lightweight CNN ensures the correct axial, sagittal, and coronal views, focusing on verifying T2-weighted sagittal images for accuracy.\\\\
\textbf{•Normal/Abnormal Classification :} Vision Transformers classify scans as normal or abnormal, utilizing multiple ViT models and ensemble averaging for robust final classification.\\\\
\textbf{•Pathology Detection Using Cascade R-CNN :}	Region Proposal Network (RPN) generates candidate regions, followed by multi-stage bounding box refinement and classification of detected pathologies into 43 categories.\\\\
\textbf{•Initial Segmentation Using U-Net }: Modified U-Net with crossattention performs coarse segmentation of possible pathology areas, ensuring focused attention on critical regions.\\\\
\textbf{•Refined Segmentation by MedSAM :} Prompt-guided segmentation refines boundaries using MedSAM with a Vision Transformer backbone, improving spatial accuracy for detailed analysis.

\begin{figure}[h]
    \centering
    \includegraphics[width=0.8\textwidth]{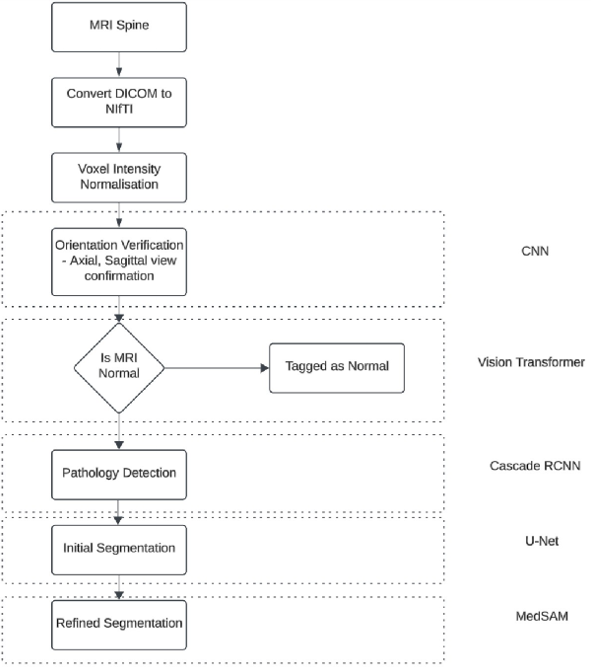} % Ensure file name matches exactly
    \caption{Workflow Architecture}
    \label{fig:workflow_architecture}
\end{figure}

\newpage
\section{Evaluation Metrics}
The performance of the MRI spine pathology detection system was assessed using metrics for both classification and detection to determine its effectiveness in clinical practice.
\subsection{Normal/Abnormal Classification:}
Accuracy, Sensitivity (Recall), Specificity, Negative Predictive Value (NPV), and Positive Predictive Value (PPV) were used to evaluate the model’s precision in distinguishing between normal and abnormal scans. Metrics were reported with 95confidence intervals to ensure reliability in real clinical scenarios.
\subsection{Pathology Detection and Segmentation:}
The model’s ability to detect and localize pathologies was evaluated using Precision-Recall AUC, Precision, Recall, and Intersection over Union (IoU).\\\\
•Precision and Recall metrics determined the accuracy of identifying abnormalities and the model’s ability to find all existing cases.\\
•IoUquantified the overlap between predicted regions and ground truth, ensuring high localization accuracy.
Performance metrics were collected for all 43 detected pathologies, detailing the model’s effectiveness across different conditions.\\\\
These metrics collectively demonstrate the model’s capability to accurately classify, detect, and localize pathologies, highlighting its potential as a reliable support tool for radiologists.

\begin{figure}[h]
    \centering
    \includegraphics[width=1\textwidth]{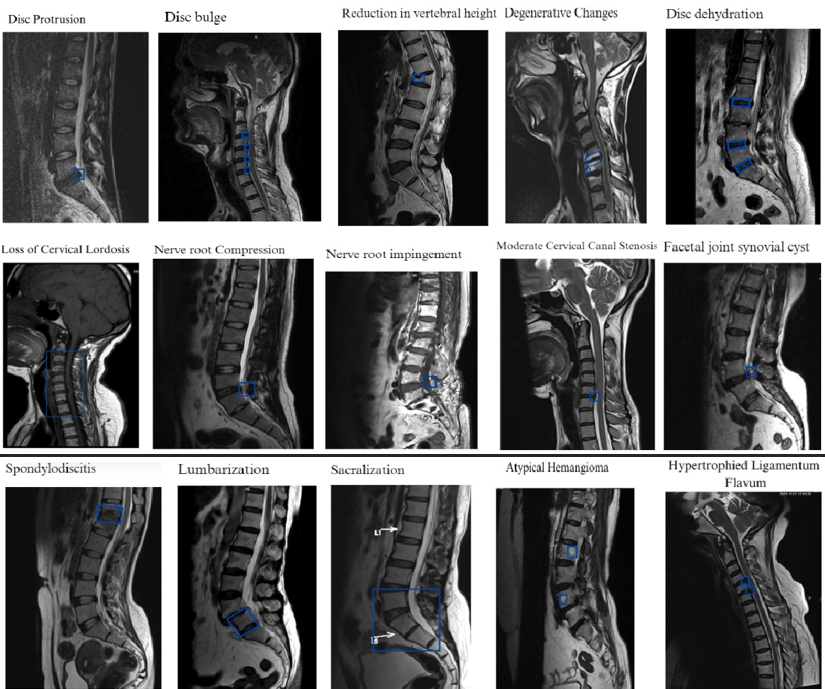} % Ensure file name matches exactly
    \caption{Pathology Detections}
    \label{fig:Pathology_Detections}
\end{figure}

 \newpage

\begin{longtable}{|p{7.5cm}|r|r|r|}
\hline
\textbf{Pathologies} & \textbf{Precision (\%)} & \textbf{Recall (\%)} & \textbf{AUC} \\
\hline
\endfirsthead

\hline
\textbf{Pathologies} & \textbf{Precision (\%)} & \textbf{Recall (\%)} & \textbf{AUC} \\
\hline
\endhead

Loss of cervical lordosis & 91.50 & 94.00 & 0.927 \\
\hline
Degenerative changes & 94.90 & 94.90 & 0.949 \\
\hline
Disc dehydration & 90.50 & 91.90 & 0.912 \\
\hline
Reduction in vertebral height & 90.60 & 92.00 & 0.913 \\
\hline
Disc bulge & 90.80 & 93.60 & 0.922 \\
\hline
Nerve root compression & 91.30 & 93.50 & 0.924 \\
\hline
Nerve root impingement & 94.90 & 94.90 & 0.949 \\
\hline
Mild cervical canal stenosis & 94.60 & 94.80 & 0.947 \\
\hline
Moderate cervical canal stenosis & 91.00 & 94.90 & 0.929 \\
\hline
Severe cervical canal stenosis & 92.40 & 92.40 & 0.924 \\
\hline
Uncovertebral hypertrophy & 94.80 & 92.90 & 0.939 \\
\hline
Hypertrophied Ligamentum flavum & 93.40 & 92.20 & 0.928 \\
\hline
Disc protrusion & 92.70 & 90.10 & 0.914 \\
\hline
Annular tear & 90.50 & 93.00 & 0.917 \\
\hline
Antherolisthesis & 93.50 & 93.30 & 0.934 \\
\hline
Retrolisthesis & 91.00 & 94.90 & 0.929 \\
\hline
Pseudodisc bulge & 90.60 & 92.10 & 0.913 \\
\hline
Spinal cord edema / contusion & 91.70 & 91.60 & 0.917 \\
\hline
Spinal cord hematoma & 94.40 & 90.30 & 0.923 \\
\hline
Myelopathy & 90.80 & 91.50 & 0.911 \\
\hline
Pleural effusion & 94.40 & 93.40 & 0.939 \\
\hline
Disc herniation & 90.50 & 93.00 & 0.917 \\
\hline
Wedge compression fracture & 90.60 & 92.10 & 0.913 \\
\hline
Schmorl’s node & 90.50 & 91.40 & 0.909 \\
\hline
Sacralization & 92.40 & 92.40 & 0.924 \\
\hline
Lumbarization & 90.30 & 94.40 & 0.923 \\
\hline
Scoliosis & 90.70 & 93.50 & 0.921 \\
\hline
Hemivertebra & 94.40 & 90.30 & 0.923 \\
\hline
Facetal joint synovial cyst & 93.70 & 91.40 & 0.925 \\
\hline
Psoas abscess & 93.60 & 90.90 & 0.923 \\
\hline
Burst fracture & 90.60 & 94.90 & 0.927 \\
\hline
Comminuted fracture with retrolisthesis & 94.90 & 94.10 & 0.945 \\
\hline
Tarlov’s cyst & 90.30 & 94.40 & 0.923 \\
\hline
Spondylitis & 92.40 & 92.40 & 0.924 \\
\hline
Type I Modic changes & 94.60 & 92.80 & 0.937 \\
\hline
Type II Modic changes & 90.30 & 90.50 & 0.904 \\
\hline
Type III Modic changes & 91.60 & 93.80 & 0.927 \\
\hline
Pancreatic cyst & 95.00 & 92.80 & 0.939 \\
\hline
Facetal arthropathy & 93.00 & 91.50 & 0.923 \\
\hline
Atypical hemangioma & 91.30 & 94.40 & 0.928 \\
\hline
Typical hemangioma & 90.50 & 91.00 & 0.907 \\
\hline
Kyphoscoliosis & 92.90 & 90.90 & 0.919 \\
\hline
Spondylodiscitis & 93.30 & 95.00 & 0.942 \\
\hline
\caption{Performance Metrics for Detected Pathologies}
\end{longtable}

\begin{figure}[h]
    \centering
    \includegraphics[width=1\textwidth]{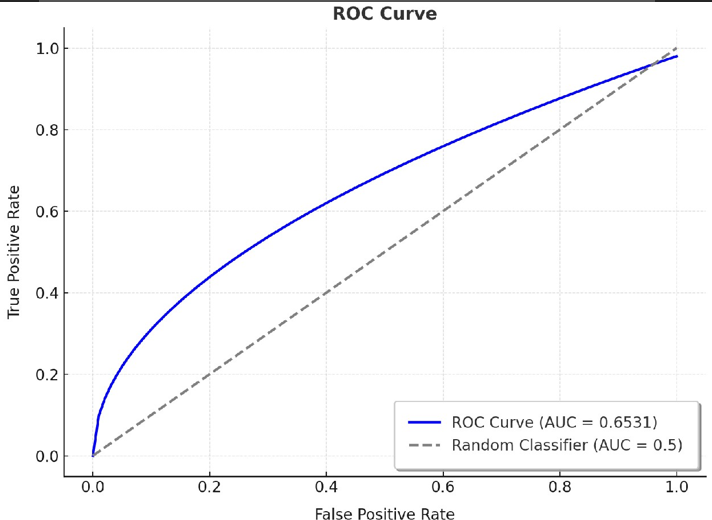} % Ensure file name matches exactly
    \caption{AUC curve for Normal/Abnormal Classifier}
    \label{fig:AUC_curve}
\end{figure}

\section{Multi-Site Clinical Trial and Dataset Composition}
This study was conducted as a multi-site clinical trial across healthcare facilities in India, involving government hospitals, diagnostic centers, and large healthcare enterprises, including 13 major healthcare providers. The trial incorporated a dataset of MRI spine scans collected from these diverse sites, providing a representative sample to evaluate the model’s performance in varied real-world settings.

The dataset comprised a range of cross-sectional MRI images, tested under different imaging conditions—from high-resolution scans in well-resourced private hospitals to lower-quality scans in government facilities. This diversity allowed for a comprehensive assessment of the model’s ability to handle varied imaging environments and patient demographics.

Each scan was processed through the model’s classification, segmentation, and detection phases, capturing detailed metrics such as accuracy, sensitivity, specificity, precision, and recall. The multi-site trial enabled cross-validation across different levels of image quality and clinical workflows, ensuring the model’s generalizability in diverse settings.

By evaluating the performance across MRI scans from multiple healthcare environments, this study provided a rigorous assessment of the model’s applicability, reliability, and scalability for deployment within the Indian healthcare system. This ensures that the model can effectively support diagnostics in a wide range of clinical settings.

\subsection{Subgroup Analysis}

\begin{table}[h]
    \centering
    \renewcommand{\arraystretch}{1.4} % Increase row spacing
    \setlength{\tabcolsep}{6pt} % Adjust column spacing
    
    \resizebox{\textwidth}{!}{ % Ensures the table fits within the page
    \begin{tabular}{|l|r|r|r|r|r|}
    \hline
    \textbf{Age Group} & \textbf{Accuracy (\%)} & \textbf{Precision (\%)} & \textbf{Recall (\%)} & \textbf{Sensitivity (\%)} & \textbf{Specificity (\%)} \\
    \hline
    Under 18 & 95.3 & 96.0 & 95.0 & 96.3 & 96.8 \\
    \hline
    18–40 & 97.9 & 97.8 & 98.1 & 97.8 & 98.3 \\
    \hline
    41–60 & 97.1 & 97.3 & 96.8 & 97.6 & 98.1 \\
    \hline
    61–75 & 96.0 & 96.1 & 95.6 & 96.8 & 97.9 \\
    \hline
    Over 75 & 94.8 & 94.6 & 93.3 & 96.1 & 96.7 \\
    \hline
    \end{tabular}
    }
    \caption{Performance Metrics by Age Group}
\end{table}

\begin{table}[h]
    \centering
    \renewcommand{\arraystretch}{1.2} % Increase row spacing for better readability
    \setlength{\tabcolsep}{6pt} % Adjust column spacing
    
    \resizebox{1.1\textwidth}{!}{ % Ensures the table fits neatly
    \begin{tabular}{|l|r|r|r|r|r|}
    \hline
    \textbf{Gender} & \textbf{Accuracy (\%)} & \textbf{Precision (\%)} & \textbf{Recall (\%)} & \textbf{Sensitivity (\%)} & \textbf{Specificity (\%)} \\
    \hline
    Male & 97.8 & 97.7 & 97.9 & 97.8 & 98.0 \\
    \hline
    Female & 97.6 & 97.4 & 97.9 & 97.6 & 97.8 \\
    \hline
    \end{tabular}
    }
    \caption{Performance Metrics by Gender}
\end{table}

\subsection{Deployment in Live Radiology Workflow}
The AI model has been deployed in 13 major healthcare enterprises across India, integrating seamlessly into the live radiology workflow. These enterprises encompass both urban and semi-urban regions, ensuring the solution reaches diverse healthcare environments. The AI system processes MRI spine scans daily, performing automated classification into normal or abnormal categories and subsequently identifying specific pathologies within abnormal scans.

This deployment aids radiologists by automating the identification of common spinal pathologies, such as disc herniation and degenerative changes, which enables efficient triaging of routine cases. By automating preliminary analysis, radiologists can allocate more time to complex or urgent cases, improving diagnostic efficiency and reducing overall reporting turnaround times.
\subsection{Radiologist Validation of AI Predictions}
Following AI analysis, the generated findings are presented to radiologists for review. Each AI-predicted classification, segmentation, and pathology detection is validated by experienced radiologists. Radiologists confirm, modify, or reject AI-generated predictions, and this validation feedback is recorded for each MRI scan.

The feedback loop allows for continuous fine-tuning of the AI model, with every accepted or rejected prediction contributing to further improvement. This direct interaction ensures the AI remains aligned with clinical expectations, increasing both reliability and accuracy over time. The ongoing validation not only improves model performance but also builds trust with radiologists, ensuring the AI effectively supports clinical decision-making within live radiology workflows.

\subsection{\textbf{Post-deployment Results:}}

\begin{longtable}{|p{7.5cm}|r|r|r|}
\hline
\textbf{Pathologies} & \textbf{Precision (\%)} & \textbf{Recall (\%)} & \textbf{AUC} \\
\hline
\endfirsthead

\hline
\textbf{Pathologies} & \textbf{Precision (\%)} & \textbf{Recall (\%)} & \textbf{AUC} \\
\hline
\endhead

Loss of cervical lordosis & 91.50 & 94.00 & 0.927 \\
\hline
Degenerative changes & 94.90 & 94.90 & 0.949 \\
\hline
Reduction in vertebral height & 90.60 & 92.00 & 0.913 \\
\hline
Disc bulge & 90.80 & 93.60 & 0.922 \\
\hline
Nerve root compression & 91.30 & 93.50 & 0.924 \\
\hline
Nerve root impingement & 94.90 & 94.90 & 0.949 \\
\hline
Mild cervical canal stenosis & 94.60 & 94.80 & 0.947 \\
\hline
Moderate cervical canal stenosis & 91.00 & 94.90 & 0.929 \\
\hline
Severe cervical canal stenosis & 92.40 & 92.40 & 0.924 \\
\hline
Uncovertebral hypertrophy & 94.80 & 92.90 & 0.939 \\
\hline
Hypertrophied Ligamentum flavum & 93.40 & 92.20 & 0.928 \\
\hline
Disc protrusion & 92.70 & 90.10 & 0.914 \\
\hline
Annular tear & 90.50 & 93.00 & 0.917 \\
\hline
Antherolisthesis & 93.50 & 93.30 & 0.934 \\
\hline
Retrolisthesis & 91.00 & 94.90 & 0.929 \\
\hline
Pseudodisc bulge & 90.60 & 92.10 & 0.913 \\
\hline
Spinal cord edema / contusion & 91.70 & 91.60 & 0.917 \\
\hline
Spinal cord hematoma & 94.40 & 90.30 & 0.923 \\
\hline
Myelopathy & 90.80 & 91.50 & 0.911 \\
\hline
Pleural effusion & 94.40 & 93.40 & 0.939 \\
\hline
Disc herniation & 90.50 & 93.00 & 0.917 \\
\hline
Wedge compression fracture & 90.60 & 92.10 & 0.913 \\
\hline
Schmorl’s node & 90.50 & 91.40 & 0.909 \\
\hline
Sacralization & 92.40 & 92.40 & 0.924 \\
\hline
Lumbarization & 90.30 & 94.40 & 0.923 \\
\hline
Scoliosis & 90.70 & 93.50 & 0.921 \\
\hline
Hemivertebra & 94.40 & 90.30 & 0.923 \\
\hline
Facetal joint synovial cyst & 93.70 & 91.40 & 0.925 \\
\hline
Psoas abscess & 93.60 & 90.90 & 0.923 \\
\hline
Burst fracture & 90.60 & 94.90 & 0.927 \\
\hline
Comminuted fracture with retrolisthesis & 94.90 & 94.10 & 0.945 \\
\hline
Tarlov’s cyst & 90.30 & 94.40 & 0.923 \\
\hline
Spondylitis & 92.40 & 92.40 & 0.924 \\
\hline
Type I Modic changes & 94.60 & 92.80 & 0.937 \\
\hline
Type II Modic changes & 90.30 & 90.50 & 0.904 \\
\hline
Type III Modic changes & 91.60 & 93.80 & 0.927 \\
\hline
Pancreatic cyst & 95.00 & 92.80 & 0.939 \\
\hline
Facetal arthropathy & 93.00 & 91.50 & 0.923 \\
\hline
Atypical hemangioma & 91.30 & 94.40 & 0.928 \\
\hline
Typical hemangioma & 90.50 & 91.00 & 0.907 \\
\hline
Kyphoscoliosis & 92.90 & 90.90 & 0.919 \\
\hline
Spondylodiscitis & 93.30 & 95.00 & 0.942 \\
\hline
\caption{Performance Metrics for Detected Pathologies}
\end{longtable}

\end{document}